\pdfoutput=0 
\documentclass{article}
\usepackage{frascatiphys}
\usepackage{graphicx}
\begin{document}
\title{ 
Review on Cosmic-Ray Radio Detection
}
\author{
Frank G. Schr\"oder\\
{\em Institut f\"ur Kernphysik, Karlsruhe Institute of Technology (KIT), Germany} 
}
\maketitle
\baselineskip=11.6pt

\begin{abstract}
Extensive air showers still are our only access to the highest-energy particles in the universe, namely cosmic-ray nuclei with energies up to several $100\,$EeV. 
Studying open questions in cosmic-ray physics, like their yet unknown origin requires the reconstruction of the energy and mass of the primary particles from the air-shower measurements. 
Great progress has been achieved lately in the development of the radio detection technique for this purpose.
There now is a consistent picture of the mechanisms behind the radio emission, which is in agreement with measurements. 
Several second-generation, digital antenna arrays are operating in different parts of the world not only aiming at the further development of the technique, but also contributing to cosmic-ray physics at energies above $100\,$PeV. Recently it has been demonstrated experimentally that radio detection can compete in precision with established techniques for air showers, like the measurement of secondary particles on ground, or fluorescence and Cherenkov light emitted by air showers. 
Consequently, cosmic-ray observatories can benefit from radio extensions to maximize their total measurement accuracy. 
\end{abstract}
\baselineskip=14pt

\section{Introduction}
Radio detection of particle cascades is one of several measurement techniques for astroparticle physics in the energy range above $10^{16}\,$eV\cite{HuegeReview2016, SchroederReview2016}. 
These cascades are primarily cosmic-ray air showers, but radio detection is also used for the search for neutrino-initiated cascades in ice and in the lunar regolith\cite{BrayReview2016}. 
This proceeding focuses on the application of the radio technique on cosmic-ray physics via the measurement of air showers. 
In any case, there is no doubt that radio detection will work also in other media and a practical demonstration will be just a question of time. 
Compared to optical techniques like the detection of Cherenkov or fluorescence light, radio detection is available around the clock and not limited by light or weather conditions, except for thunderstorms directly above the radio antennas\cite{LOPESthunderstorm2011}. 
The first radio measurements of air showers took place already in the 1960's, though with limited accuracy due to the analog electronics available at that time\cite{AllanReview1971}.
Current antenna arrays have reached measurement accuracies similar to the established optical techniques in the energy range above $10^{17}\,$eV, and in a few years the SKA can achieve an even higher precision and a lower threshold around $10^{16}\,$eV\cite{SKA_ICRC2015}.

For air showers the dominant mechanism of radio emission is the deflection of the electrons and positrons in the geomagnetic field, which induces a transverse current in the shower front.
This leads to linearly-polarized radio emission whose strength increases with the local size of the geomagnetic field, and with $\sin \alpha$, i.e., the angle between the shower axis and the geomagnetic field\cite{CODALEMA_Geomagnetic}.
The strength of the geomagnetic emission also depends on the density of the medium: 
In inclined showers developing higher up in the air the emission region around the shower maximum is more extended.
Thus, there is more energy emitted in radio waves than for vertical showers\cite{GlaserShortAuthor2016}. 
In dense media like ice, to the contrary, the showers are so compact that geomagnetic emission is negligible. 

As a second mechanism the Askaryan effect contributes, i.e., radially polarized emission due to the time-variation of the electron excess in the shower front. 
The Askaryan effect has similar strength in all media, which makes it the dominant mechanism in dense media, where the geomagnetic emission is negligible.
The strength in air depends on the shower inclination and distance to the shower axis\cite{SchellartLOFARpolarization2014}.
Typically the amplitude of the Askaryan effect is only $10-20\,\%$ of the geomagnetic amplitude. 
The radio emission of air showers seems to be understood to at least this level of $10-20\,\%$, and current simulation programs for the radio emission, such as CoREAS, agree with measurements of the absolute radio amplitude to better than $20\,\%$ \cite{TunkaRex_NIM2015, LOPESimprovedCalibration2016, LOFARcalibration2015}.

\begin{figure}[t]
  \centering
  \includegraphics[width=0.99\linewidth]{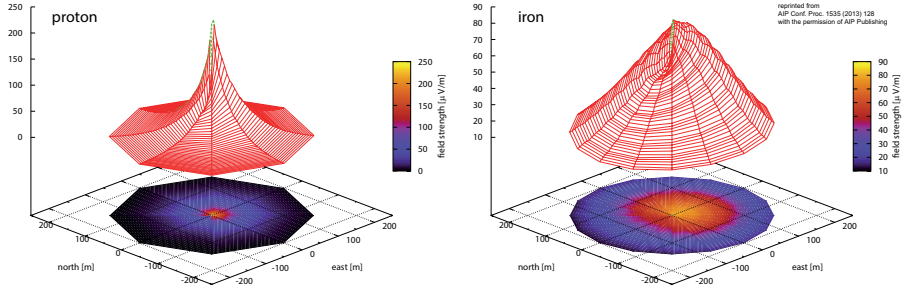}
  \caption{CoREAS simulations of the radio emission by two air showers, one initiated by a proton, and one by an iron nucleus.
  The height and the color code indicate the amplitude at ground level. 
  The steepness of the footprint depends on the distance to the shower maximum; the small asymmetry is caused by the interference of the geomagnetic and Askaryan effects\cite{HuegeCoREAS_ARENA2012}.}
  \label{fig_CoREASfootprints}
\end{figure}

For both emission mechanisms the radio emission is coherent when the wavelength is larger than the optical pathlength of the shower front. 
For the frequency band of $30-80\,$MHz chosen by many experiments, this is the case up to a few $100\,$m distance from the shower axis. 
This means that radio emission is forward beamed into a narrow cone with an opening angle of the order of $3^\circ$ (see figure \ref{fig_CoREASfootprints}).
As a consequence, antenna arrays have to be either relatively dense with antenna spacings on the order of $100\,$m, or have to target inclined showers, since the illuminated area on the ground increases with the distance to the shower maximum. 
Recently, the Auger Engineering Radio Array (AERA) has measured that for inclined showers the radio footprint has a size similar to the particle footprint of several km$^2$ at $10^{18}\,$eV \cite{AERAoverviewICRC2015}.

\begin{figure}[t]
  \centering
  \includegraphics[width=0.99\linewidth]{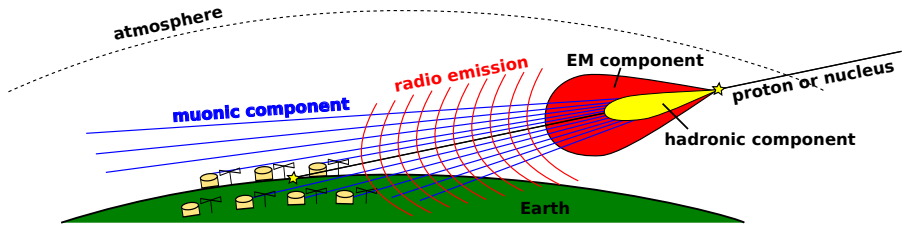}
  \caption{Sketch of an inclined air shower. The electromagnetic component is absorbed in the air and only muons can be measured at the ground - in addition to the large footprint of the radio emission by the electromagnetic component\cite{AERAinclinded_ARENA2014}.}
  \label{fig_inclinedSketch}
\end{figure}

For a given shower direction the radio amplitude is proportional to the number of electrons, and radio detection provides a calorimetric measurement of the shower energy similarly to air-fluorescence or air-Cherenkov light detection\cite{AERAenergyPRL, AERAenergyPRD}.
This makes radio detection complementary to particle detectors, which can measure only muons for inclined showers, since the electromagnetic component is absorbed in the air (see figure \ref{fig_inclinedSketch}). 
Since the radio measurement of the calorimetric shower energy in combination with the number of muons depends statistically on the mass of primary particle, this combination of radio and particle detectors is a very promising technique for future large-scale arrays\cite{Holt_TAUP}. 
Additionally the atmospheric depth of the shower maximum provides complementary information on the type of the primary particle. 
However, before applying the radio technique on a large scale for inclined showers, some investigation is still necessary on how accurately the shower energy and the position of the shower maximum can be measured. 
Nonetheless, these analysis techniques are already fairly advanced for air showers with zenith angles below $60^\circ$. 
Current antenna arrays have achieved reconstruction precisions for the energy and the shower maximum comparable to those of the leading optical techniques, i.e., about $15\,\%$ energy precision and about $20\,$g/cm$^2$ for the atmospheric depth of the shower maximum.

\section{Precision of shower parameters}
Radio detection is sensitive to the three shower parameters most important for cosmic-ray physics:
the direction of the shower axis, which is equal to the arrival direction of the primary particle; 
the shower energy, which is an estimator for the energy of the primary particle;
the atmospheric depth of the shower maximum, $X_\mathrm{max}$, which is an estimator for the mass composition of the primary cosmic rays.

The \emph{direction} can be reconstructed with a precision of better than $0.7^\circ$, as shown by LOPES featuring nanosecond-precise time calibration\cite{SchroederTimeCalibration2010} and using digital radio interferometry\cite{LOPESwavefront2014}. 
With a very dense and accurately synchronized array, such as LOFAR, a resolution of even $0.1^\circ$ might be possible\cite{CorstanjeLOFARtimeCalibration2016}, though $1^\circ$ resolution usually is sufficient since charged cosmic rays are deflected anyway by magnetic fields on their way to Earth.

\begin{figure}[t]
  \centering
  \includegraphics[width=0.46\linewidth]{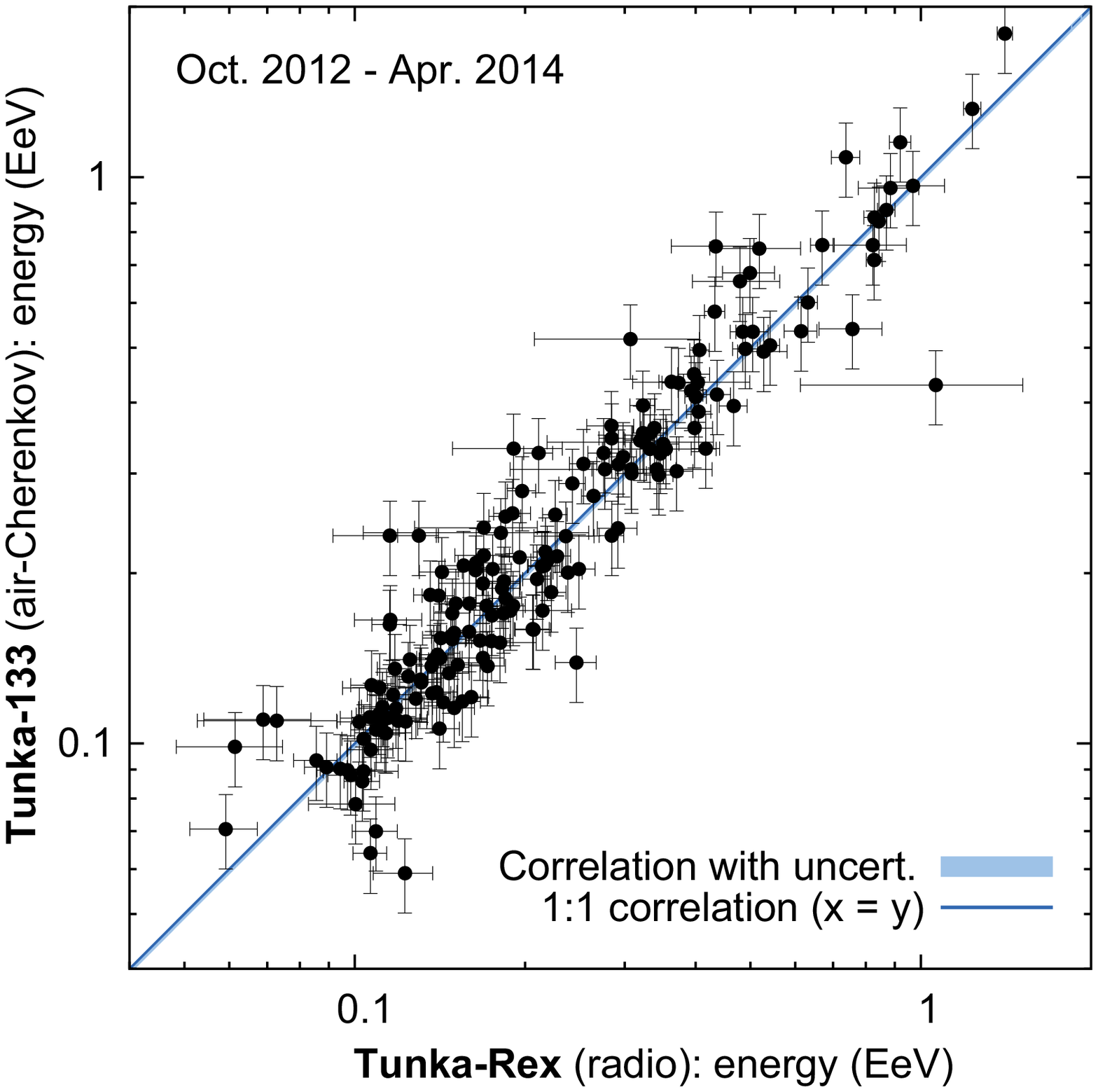}
  \hfill
  \includegraphics[width=0.50\linewidth]{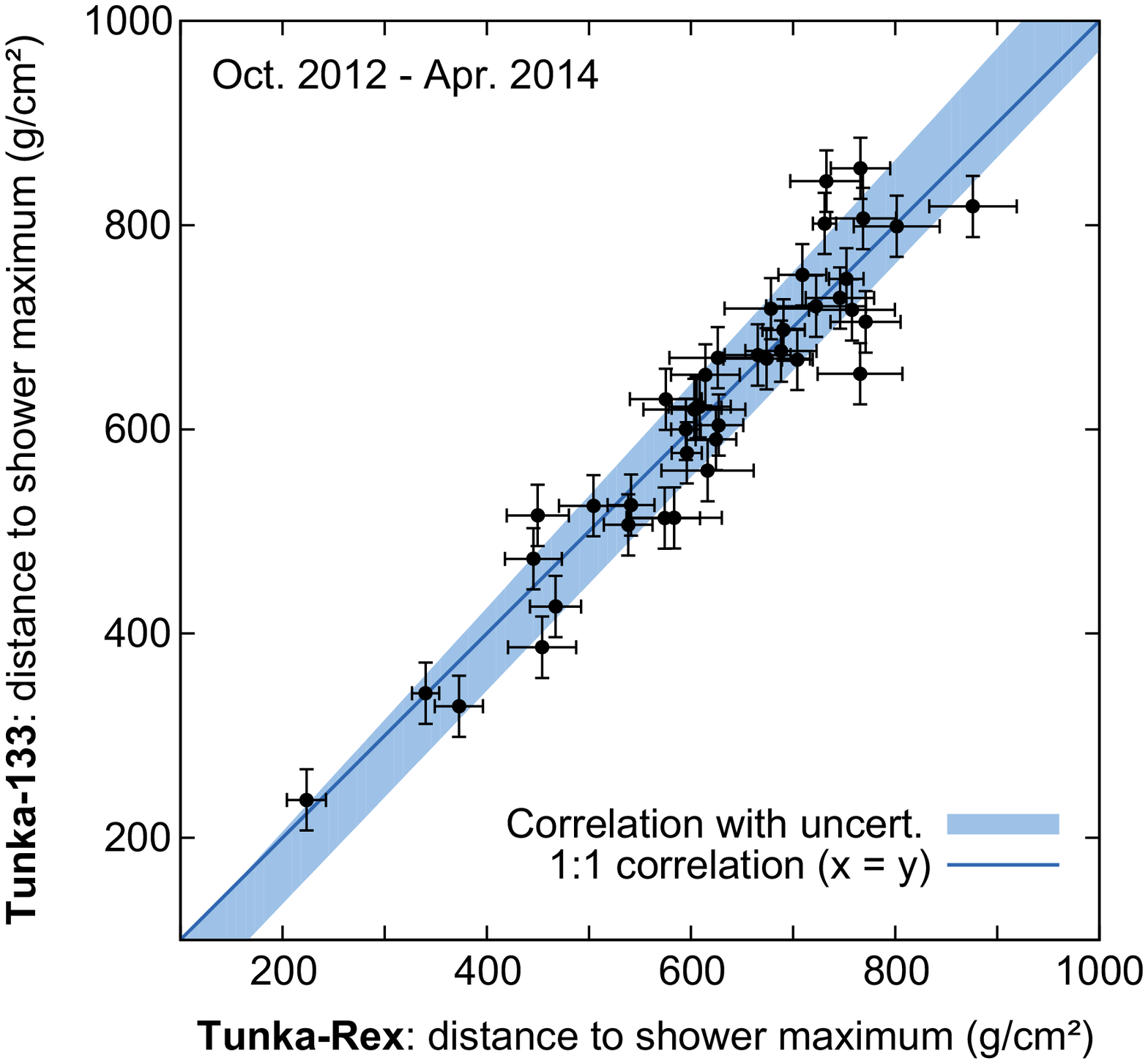}
  \caption{Radio measurements of the shower energy and of $X_\mathrm{max}$ by Tunka-Rex compared to the coincident air-Cherenkov measurements by Tunka-133\cite{TunkaRex_XmaxJCAP2016}.}
  \label{fig_tunkaEnergyAndXmax}
\end{figure}

There are two ways to reconstruct the \emph{energy} from measurements of the radio amplitude: 
First, the time and space integral of the signal power yields the total radiation energy of the shower at radio frequencies. 
This radiation energy increases quadratically with the shower energy due to the coherent nature of the radio emission. 
AERA has demonstrated a precision and a scale uncertainty for estimating the energy of the primary particle by this method of better than $20\,\%$\cite{AERAenergyPRL, AERAantennaJINST2012}. 
Second, the radio amplitude at a detector specific distance is proportional to the shower energy. 
While this method is not as universal, it might be more robust for measurements close to the detection threshold. 
The precision demonstrated by this method is similar to the first one, e.g., about $20\,\%$ for LOPES\cite{2014ApelLOPES_MassComposition}, and about $15\,\%$ for Tunka-Rex (see figure \ref{fig_tunkaEnergyAndXmax})\cite{TunkaRex_XmaxJCAP2016}.

The position of the \emph{shower maximum} is the parameter most difficult to reconstruct.
Nevertheless, there are several methods for this, since several properties of the radio signal depend on the distance to the shower maximum.
LOPES\cite{2012ApelLOPES_MTD, 2014ApelLOPES_MassComposition} and Tunka-Rex\cite{TunkaRex_XmaxJCAP2016, KostuninTheory2015} have shown that the slope of the lateral distribution can be used, and Tunka-Rex has achieved a precision of $40\,$g/cm$^2$, which is twice the value achieved by the leading air-fluorescence technique. 
Moreover the steepness of the hyperbolic radio wavefront\cite{LOPESwavefront2014} and the slope of the frequency spectrum\cite{Grebe_ARENA2012} are sensitive to $X_\mathrm{max}$, but the precision achievable under practical conditions is not yet clear. 

\begin{figure}[t]
  \centering
  \includegraphics[width=0.99\linewidth]{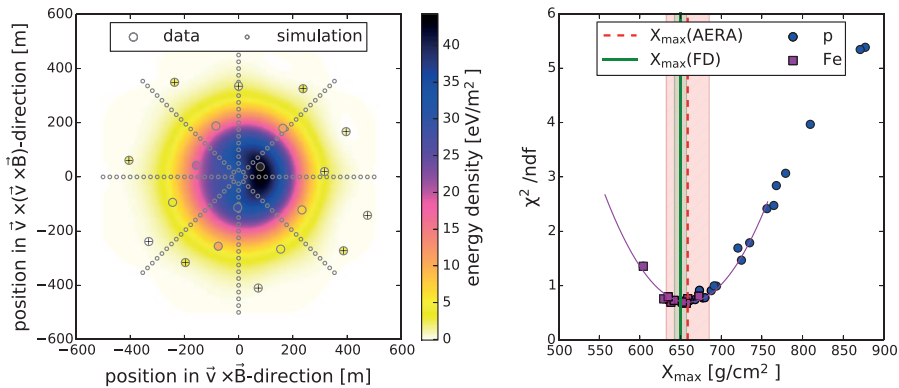}
  \caption{Top-down reconstruction method for $X_\mathrm{max}$ for an example event measured by AERA. 
  Left: the simulated radio amplitude is matched with the measurements by adjusting the amplitude scale and the core position; crosses indicate sub-threshold stations.
  Right: the $X_\mathrm{max}$ value of the best fitting simulation is assumed as the real $X_\mathrm{max}$ of the measured air showers, which for this example event is confirmed by coincident fluorescence measurements (FD)\cite{AERAoverviewICRC2015}.}
  \label{fig_AERAtopDownXmax}
\end{figure}

The most precise, but also most computationally intensive method for $X_\mathrm{max}$ is a top-down approach introduced by LOFAR\cite{BuitinkLOFAR_Xmax2014} and meanwhile also applied by AERA\cite{AERAoverviewICRC2015}. 
Several Monte Carlo simulations with different distances to the shower maximum are produced for the shower geometry of an individual event. 
Then, the simulated amplitude is compared to the amplitude measured at the various antenna stations to check for which $X_\mathrm{max}$ the simulations fit best (see figure \ref{fig_AERAtopDownXmax}). 
Hence, the method implicitly exploits all $X_\mathrm{max}$ sensitive characteristics of the radio footprint, not just its slope. 
Featuring more than 100 antennas per event LOFAR demonstrated a precision of better than $20\,$g/cm$^2$ for $X_\mathrm{max}$ by this method\cite{LOFAR_Nature2016}.
This precision is already similar to that of air-fluorescence measurements and might be further improved by including the information of the wavefront, the frequency spectrum, and the polarization.

\section{Conclusion}
Due to significant progress in the development of the radio technique and in the understanding of the emission mechanism, radio measurements can now compete in precision with optical techniques for air showers, and this around the clock. 
Air-shower arrays made of particle detectors can especially profit from a radio extension providing more accurate information on the shower energy and mass composition.
Moreover, there are at least two further applications of the radio technique for cosmic-ray science. 
By focusing on inclined showers huge radio arrays covering more than $100,000\,$km$^2$, such as GRAND\cite{GRAND_ICRC2015}, could acquire significant exposure for the highest-energy extragalactic cosmic rays at several $100\,$EeV, and simultaneously the search for ultra-high-energy neutrinos at EeV energies. 
Complementary to this, radio detection can increase our knowledge on the transition from galactic to extragalactic cosmic rays, assumed in the energy range above $10^{17}\,$eV \cite{KG_lightAnkle}. 
With several $10,000$ antennas, i.e., a number similar to GRAND, but inside one square kilometer, the low-frequency core of the SKA\cite{SKA_ICRC2015} will measure air showers much more precisely than possible by the optical technique today.

\section*{Acknowledgements}
Thanks go to my colleagues at KIT, to the LOPES, Pierre Auger and Tunka-Rex Collaborations for fruitful discussions, and to DFG for grant Schr 1480/1-1. 

%

%
\end{document}